\documentclass[aps,prl,twocolumn,float,showpacs]{revtex4}

\usepackage{amsmath,bm,epsfig}
 \usepackage{hyperref}

\usepackage{pstricks}
\usepackage{pst-node}
\usepackage[ansinew]{inputenc}
\usepackage{amssymb,amsmath}

\def\a{\alpha}

\def\o{\omega}

\def\be{\begin{equation}}

\def\ee{\end{equation}}

\def\bea {\begin{eqnarray}}      \def\eea {\end{eqnarray}}

\def\bean{\begin{eqnarray*}}    \def\eean{\end{eqnarray*}}

\def\<{\langle} \def\({\left(}  \def\>{\rangle} \def\){\right)}

\begin{document}

\title{Effect of the dynamical phases on the nonlinear amplitudes' evolution}

\author{Miguel D. Bustamante$^{\dag}$  and Elena Kartashova}
 \email{lena@risc.uni-linz.ac.at}
\affiliation{$^\dag$ School of Mathematical Sciences, University
College Dublin, Belfield, Dublin 4, Ireland, EU \\
$^*$ RISC, J.Kepler University, Linz 4040, Austria,EU}

\begin{abstract}
In this Letter we show how the nonlinear evolution of a resonant triad depends on
the special combination of the modes' phases chosen according to the resonance
conditions. This phase combination is called dynamical phase. Its evolution is
studied for two integrable cases: a triad and a cluster formed by two connected
triads, using a numerical method which is fully validated by monitoring the
conserved quantities known analytically. We show that dynamical phases, usually
regarded as equal to zero or constants, play a substantial role in the dynamics of
the clusters. Indeed, some effects are (i) to diminish the period of energy
exchange $\tau$ within a cluster by 20$\%$ and more; (ii) to diminish, at time
scale $\tau$, the variability of wave energies by 25$\%$ and more; (iii) to
generate a new time scale, $T >> \tau$, in which we observe considerable energy
exchange within a cluster, as well as a periodic behaviour (with period $T$) in
the variability of modes' energies. These findings can be applied, for example, to
the control of energy input, exchange and output in Tokamaks; for explanation of
some experimental results; to guide and improve the performance of experiments; to
interpret the results of numerical simulations, etc.
\end{abstract}






\maketitle

\noindent \textbf{1. Introduction.} Nonlinear resonances are ubiquitous in
physics. Euler equations, regarded with various boundary conditions and specific
values of some parameters, describe an enormous number of nonlinear dispersive
wave systems (capillary waves, surface water waves, atmospheric planetary waves,
drift waves in plasma, etc.) all possessing nonlinear resonances \cite{zakh92}.
Nonlinear resonances appear in  mechanics \cite{Mech}, astronomy \cite{Astr},
medicine \cite{cancer}, etc., etc.
\begin{figure*}
\begin{center}
\includegraphics[width=3cm,height=5.5cm,angle=270]{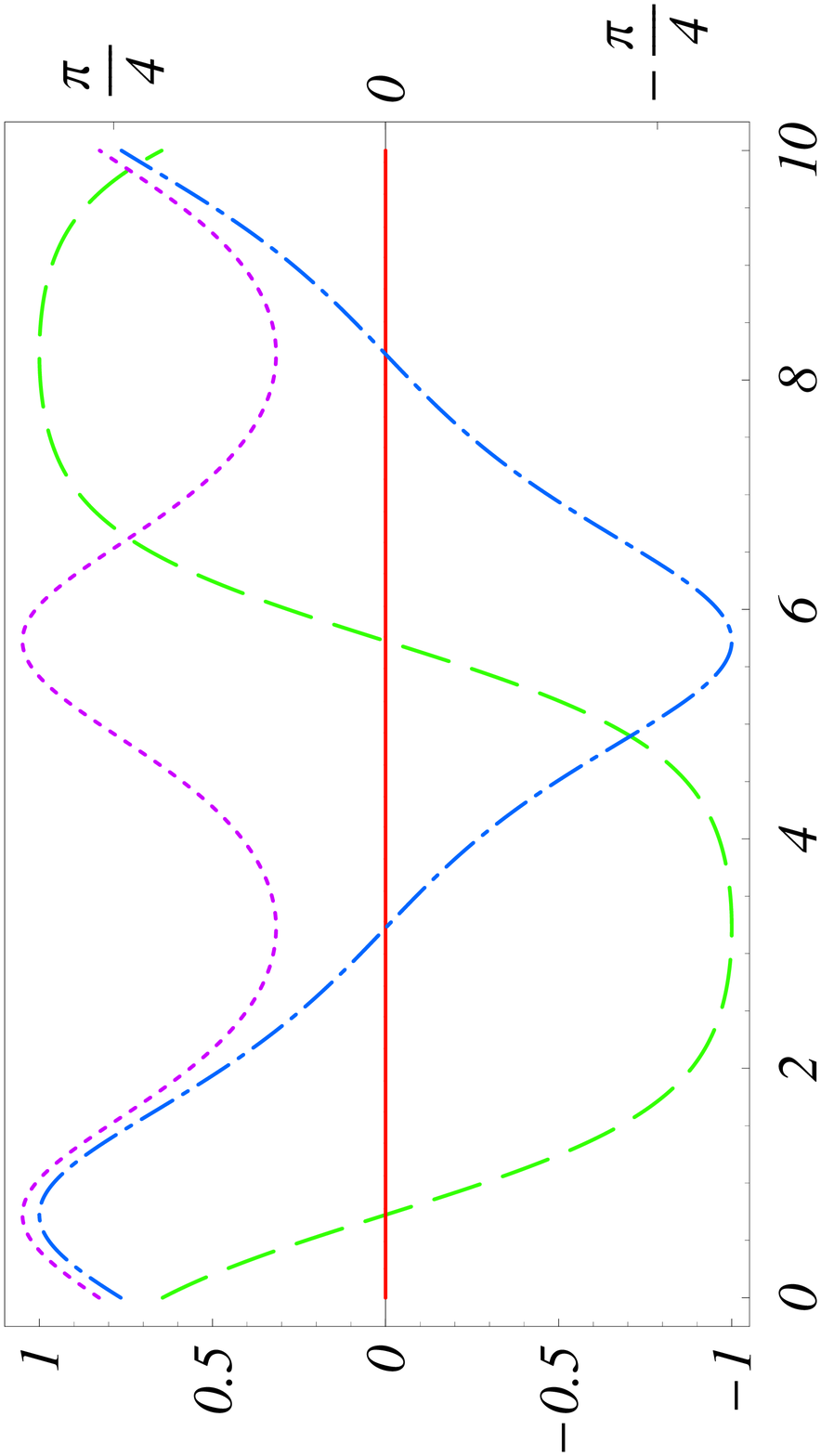}
\includegraphics[width=3cm,height=5.5cm,angle=270]{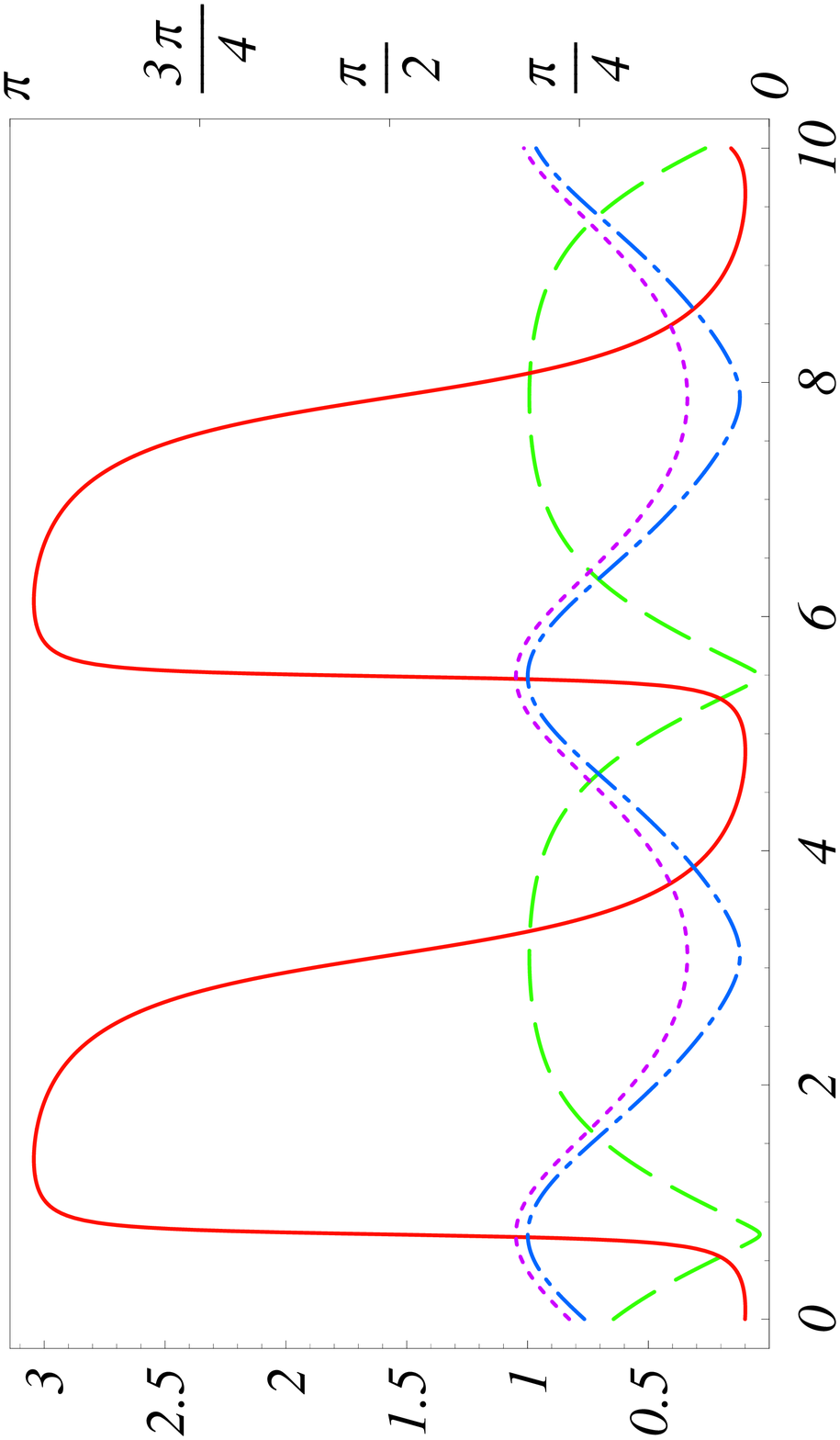}
\includegraphics[width=3cm,height=5.5cm,angle=270]{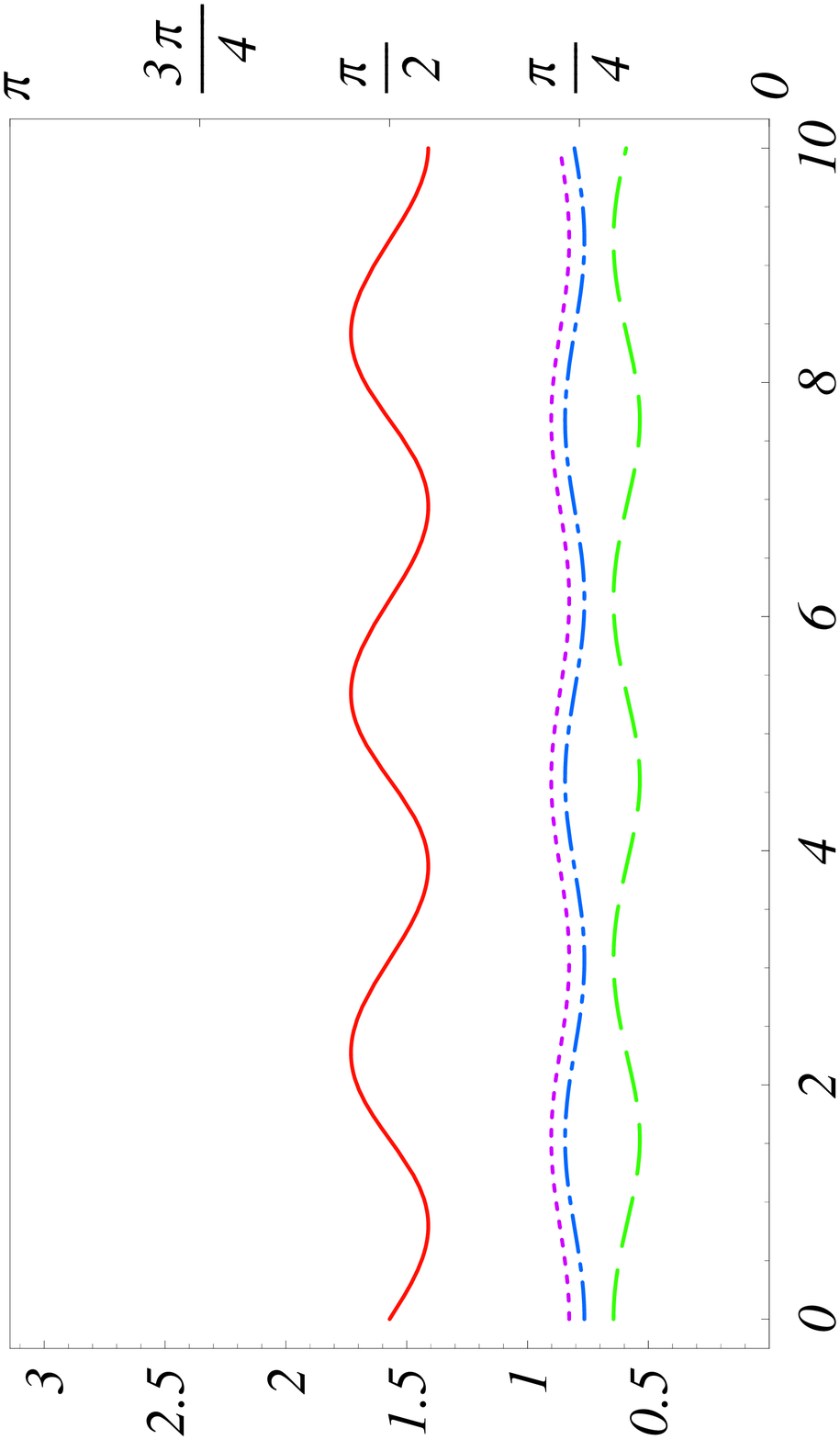}
\includegraphics[width=3cm,height=5.5cm,angle=270]{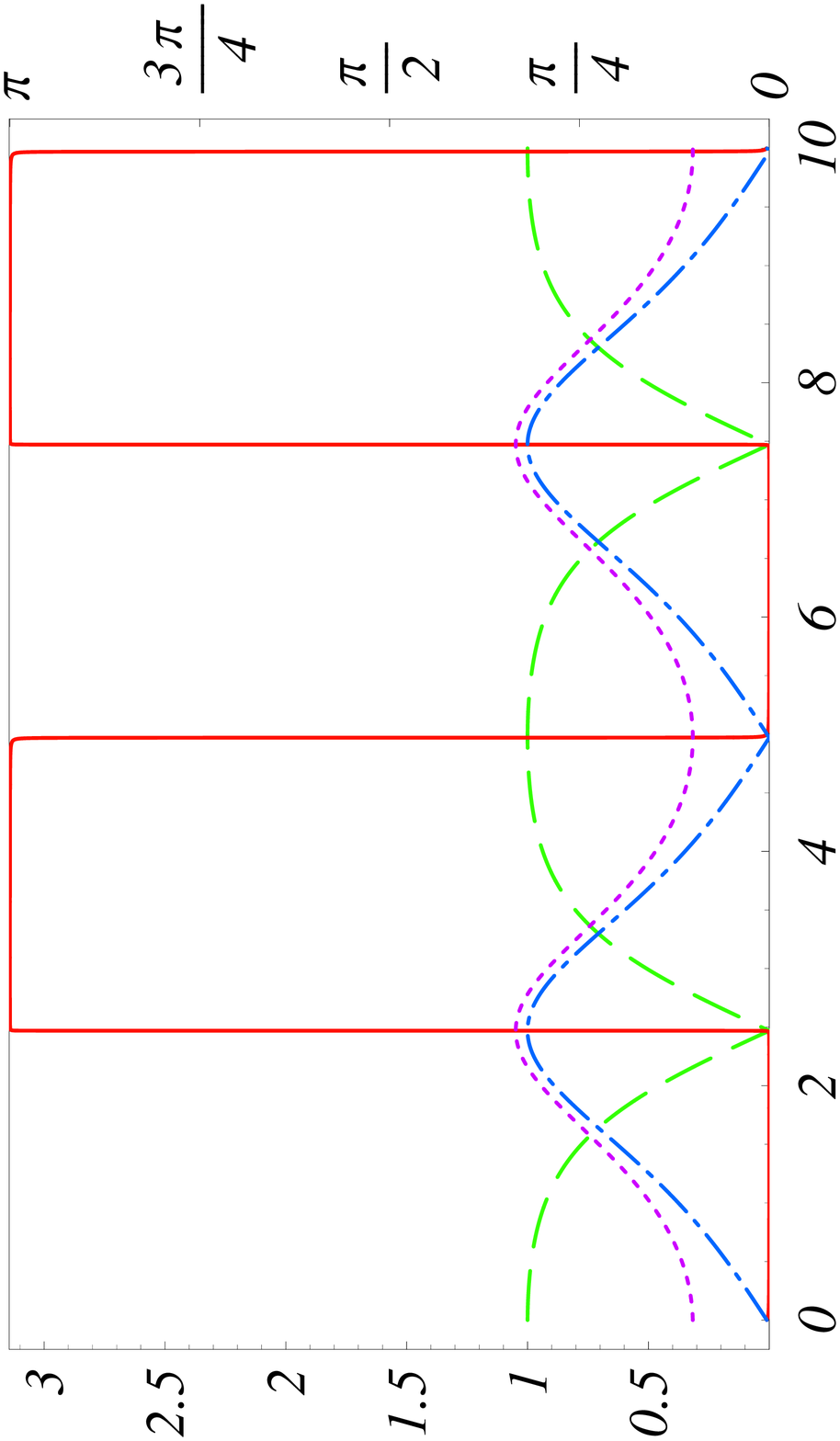}
\includegraphics[width=3cm,height=5.5cm,angle=270]{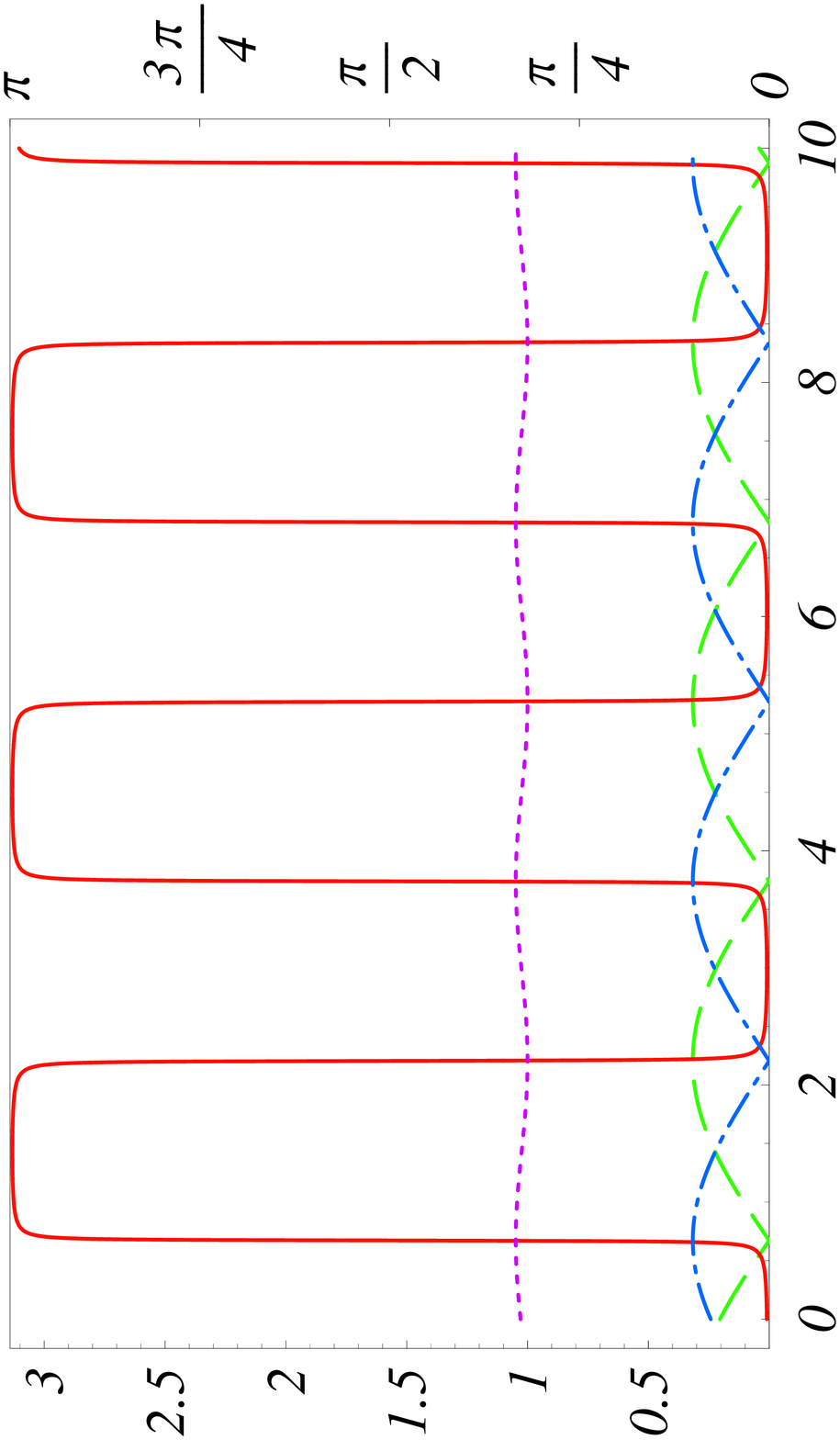}
\includegraphics[width=3cm,height=5.5cm,angle=270]{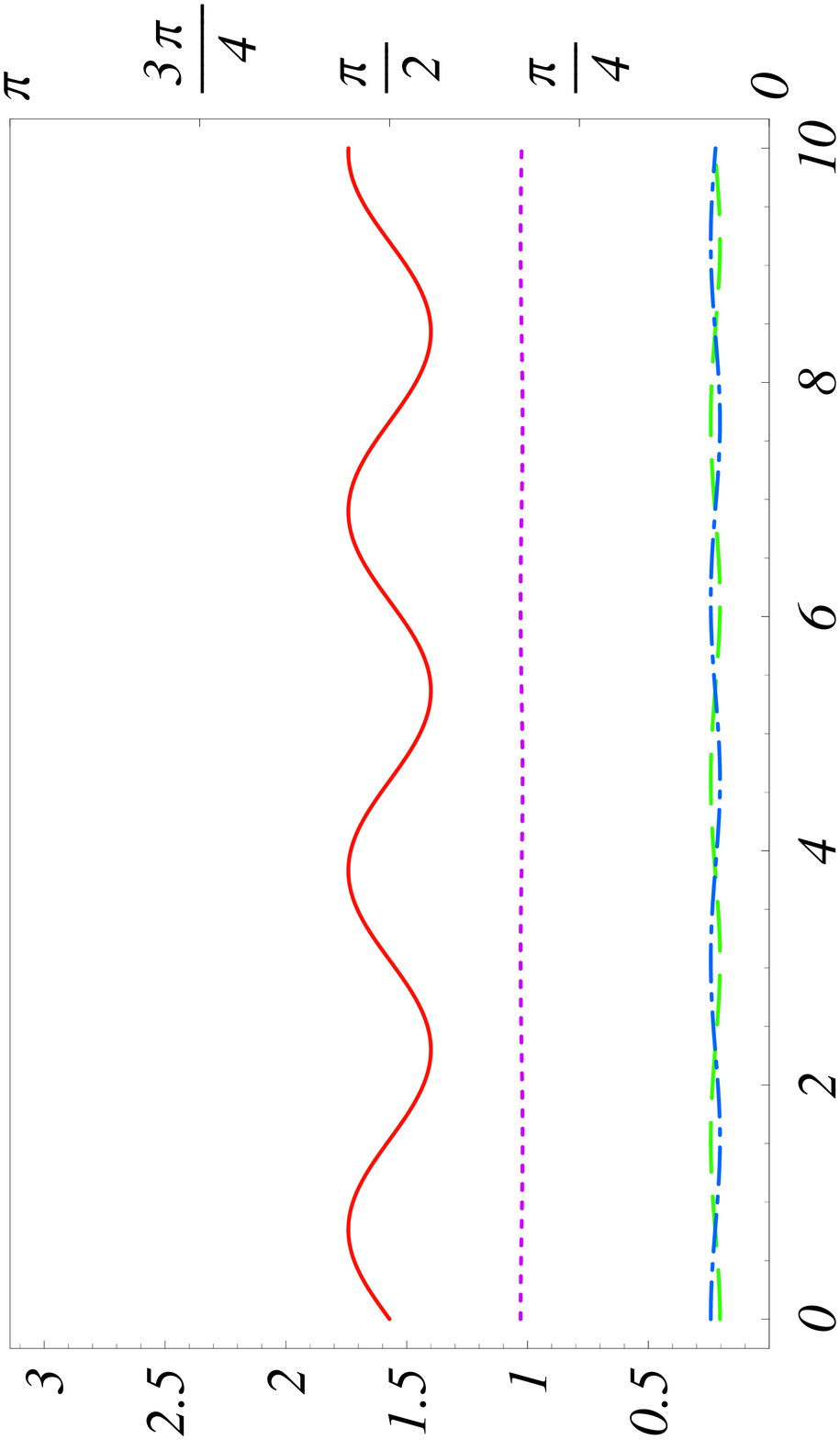}
\end{center}
\caption{\label{1} Color online. Plots of the modes' amplitudes and dynamical
phase as functions of time, for a triad with $Z=1$. For each frame, $\varphi(t)$
is (red) solid, $C_1(t)$ is (purple) dotted, $C_2(t)$ is (blue) dash-dotted,
$C_3(t)$ is (green) dashed. \textbf{Upper panel}. Initial condition $\a=0.7$ and
conserved quantities $I_{23}=1$, $I_{13}=1.1$ for all frames. Initial conditions
for the phase and corresponding cubic conserved quantity: \textbf{Left}:
$\varphi=0, I_T=0.$ \textbf{Middle}: $\varphi=0.1, I_T=0.041.$ \textbf{Right}:
$\varphi=\pi/2, I_T=0.408.$ \textbf{Lower panel}: Conserved quantity $I_{13}=1.1$
for all frames. Initial conditions and corresponding conserved quantities:
\textbf{Left}: $\a = 1.56, \varphi= 0.01$ and $I_{23}=1, I_T = 3.2 \times
10^{-5}.$ \textbf{Middle}: $\a=0.7, \varphi=0.01$ and $I_{23}=0.1, I_T = 5.1
\times 10^{-4}$. \textbf{Right}: $\a=0.7, \varphi=\pi/2$ and $ I_{23}=0.1, I_T =
5.1 \times 10^{-2}$. Here, horizontal axe denotes non-dimensional time; vertical
left and right axes denote amplitude and phase correspondingly.}
\end{figure*}
In this Letter we will regard the simplest nonlinear resonant systems
corresponding to the 3-wave resonance conditions. Examples of these nonlinear
resonant systems are, in order of simplicity, triads (which are integrable), and
small groups of connected triads which are known to be important for various
physical applications (large-scale motions in the Earth's atmosphere \cite{KL-06},
laboratory experiments with gravity-capillary waves \cite{CHS96}, etc.). Dynamical
system for a triad will be regarded in
the standard Manley-Rowe form:%
 \be \label{dyn3waves}
\dot{B}_1=   Z B_2^*B_3,\quad
\dot{B}_2= Z B_1^* B_3, \quad \dot{B}_3= -  Z B_1 B_2, %
\ee 
where $(B_1,B_2,B_3)$ are complex amplitudes of 3 resonantly interacting modes
$B_j \exp i(\mathbf{k}_j \cdot \mathbf{x} -\omega ({\bf k}_j) t )$, while the
corresponding resonance conditions are
 \be
\label{res} \omega ({\bf k}_1) + \omega ({\bf k}_2)- \omega ({\bf k}_{3}) = 0\,,
\quad {\bf k}_1 + {\bf k}_2 -{\bf k}_{3} = 0,
\ee %
where $\o(\mathbf{k})$ is the dispersion relation and $\mathbf{k}$ is the
wavevector.

\noindent Sys. (\ref{dyn3waves}) has been studied both in its real and complex
form by numerous researchers (e.g., \cite{book-triad}, \cite{LynchHoughton04},
etc.). If regarded in the amplitude-phase representation $B_j = |B_j| \exp i
\theta_j$, Sys.(\ref{dyn3waves}) is equivalent to a system for the 3 real
amplitudes $|B_j|$ and the phase combination $\varphi = \theta_1 + \theta_2 -
\theta_3$, the individual phases $\theta_j$ being slave variables and can be
obtained by quadratures \cite{HolmLynch02}. The dynamical equation for $\varphi$
is also known (\cite{LynchHoughton04}, p.43, Eq.(28)). Still, a sort of general
misunderstanding persists, concerning the relevance of $\varphi$ for the general
dynamics of the system. It is a common belief that for an exact resonance to
occur, it is necessary that the phase $\varphi$ is either zero
(\cite{LongHigGill67}, p.132, Eq.(6.7); \cite{ped}, p.156, Eq.(3.26.19), etc.) or
constant (e.g. \cite{CHS96}).

\noindent The main goal in this Letter is to show that in the case of generic
initial conditions, the phase $\varphi$ which we call from here on \emph{dynamical
phase} affects the evolution of the amplitudes and therefore, has a direct impact
on the behaviour of any physical system governed by a triad as well as small
clusters of resonant triads. As it was shown in \cite{BK08}, some of these
clusters are described by integrable systems and for them a complete set of
conservation laws (CLs) was given explicitly, showing that dynamical phases are
relevant in the determination of these CLs. In this paper we present differential
equations for the two independent dynamical phases appearing in the
\emph{butterfly}, a resonance cluster formed by two triads connected \emph{via}
one mode. We investigate the integrable case of butterfly by solving numerically
the reduced evolutionary differential equations shown in \cite{BK08} for the
phases and amplitudes. The numerical integration is fully validated by monitoring
the conservation laws, known analytically from our previous work \cite{BK08}. We
study the effects of the phases on the modes' amplitudes, and some physical
implications are briefly discussed.

\noindent \textbf{2. Triad with complex amplitudes.} The CLs for the
Sys.(\ref{dyn3waves}) have the form \bea\label{laws3waves}
 \begin{cases}
 I_{13}= |B_1 |^2 + |B_3|^2,\quad
 I_{23}=|B_2 |^2 + |B_3|^2 \,,\\
 I_T = \operatorname{Im}(B_1 B_{2} B_{3}^*)\,,
 \end{cases}
\eea
and their knowledge is enough for the explicit solution to be constructed. Here
conserved quantities $ I_{13}$ and $ I_{23}$ are not energy and enstrophy anymore
but their linear combinations \cite{KL-06}. The analytical solution of
Sys.(\ref{dyn3waves}) can be found in \cite{LynchHoughton04}, as well as the
equations on two of the three phases $\theta_j$ (the standard  amplitude-phase
representation $B_j= C_j\exp(i \theta_j)$ is used). Below we use  slightly
different notations, introduced in \cite{BK08}, because they are more convenient
for further studies of bigger groups of connected triads. The equation for the
dynamical phase can be easily deduced
and reads as %
\be \label{DynPhaseT} \dot{\varphi}= -I_T (C_1^{-2}+C_2^{-2}-C_3^{-2}).
 \ee
Combining (\ref{laws3waves}) and (\ref{DynPhaseT}), it is easy to see that the
constraint $I_T=0$ implies that dynamical phase vanishes, i.e. $\varphi=0$.

\noindent  If we put $I_T=0,$
the solution for amplitudes takes a very simple and familiar form:%
 \bea \label{sol3wavesReal}
\begin{cases}
C_1(t) = \mathrm{dn}(\left( -t + {t_0} \right) \,z\,
    {\sqrt{{{{I} }_{13}}}},
   \frac{{{{I} }_{23}}}{{{{I} }_{13}}})\,
  {\sqrt{{{{{I} }}_{13}}}}\\
C_2(t) = \mathrm{cn}(\left( -t + {t_0} \right) \,z\,
    {\sqrt{{{{I} }_{13}}}},
   \frac{{{{I} }_{23}}}{{{{I} }_{13}}})\,
  {\sqrt{{{{{I} }}_{23}}}}\\
C_3(t) = \mathrm{sn}(\left( -t + {t_0} \right) \,z\,
    {\sqrt{{{{I} }_{13}}}},
   \frac{{{{I} }_{23}}}{{{{I} }_{13}}})\,
  {\sqrt{{{{{I} }}_{23}}}}
\end{cases}
\eea%
where $t_0$ is defined by initial conditions and can also be written out
explicitly.

\begin{figure*}
\begin{center}
\includegraphics[width=3cm,height=5.5cm,angle=270]{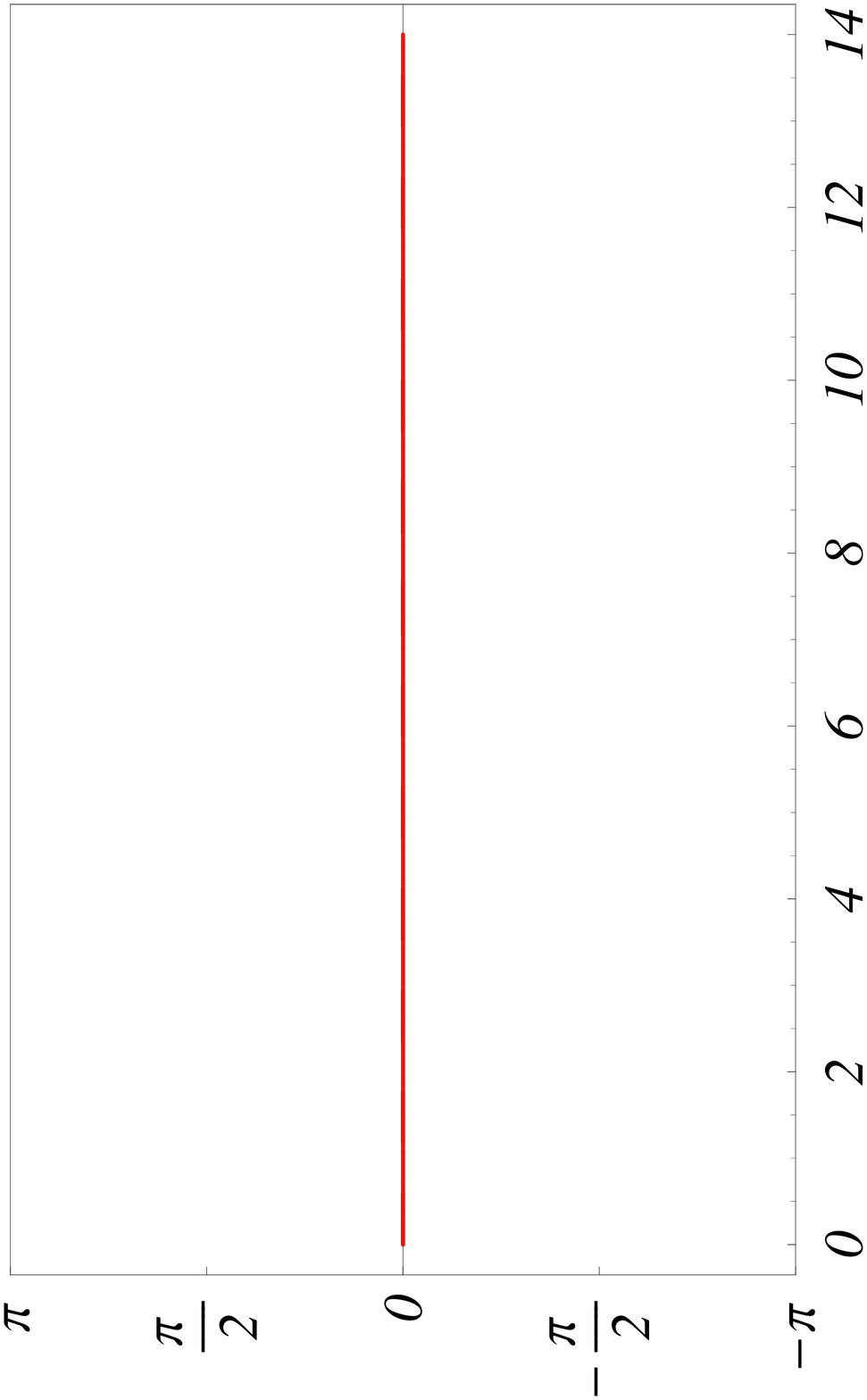}
\includegraphics[width=3cm,height=5.5cm,angle=270]{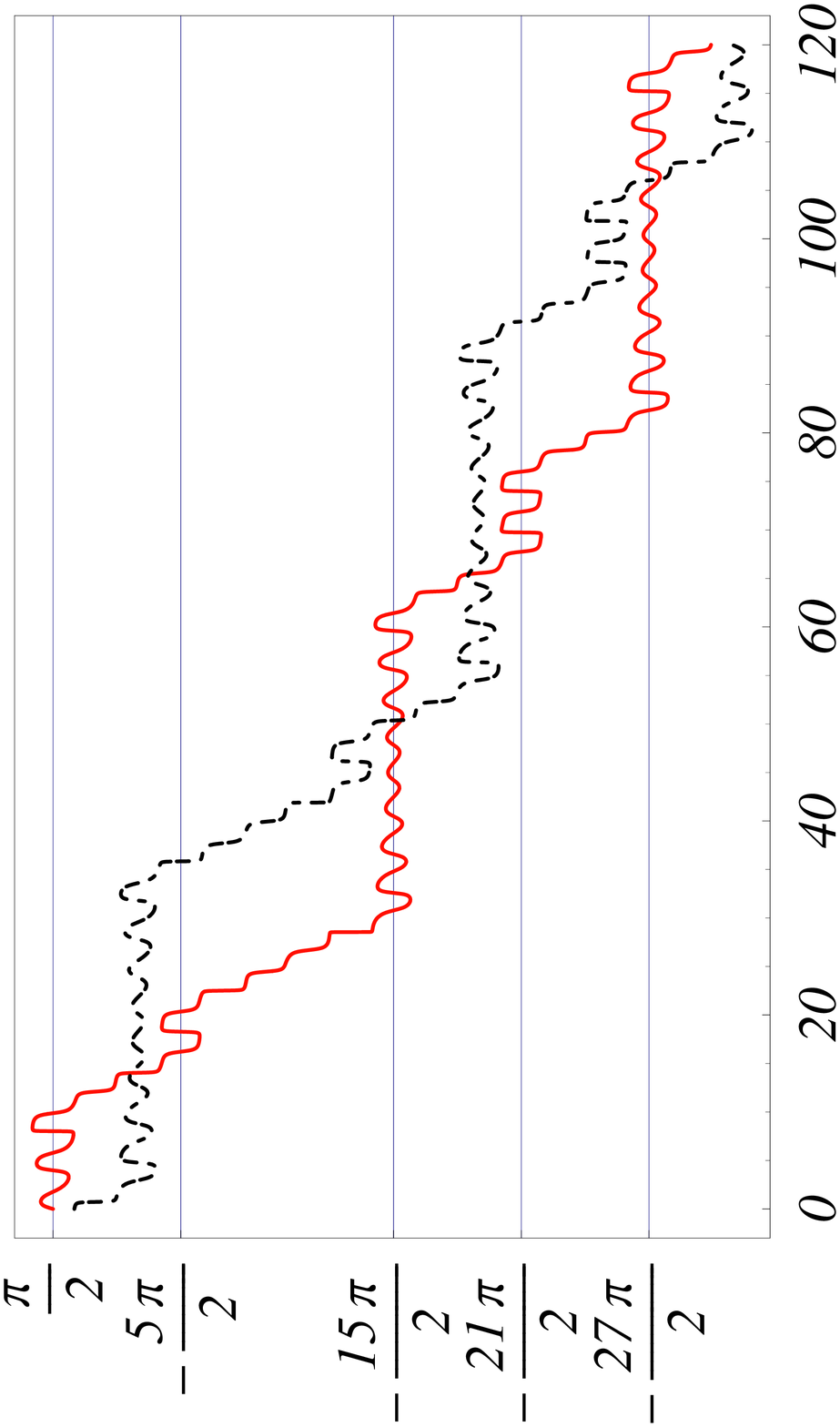}
\includegraphics[width=3cm,height=5.5cm,angle=270]{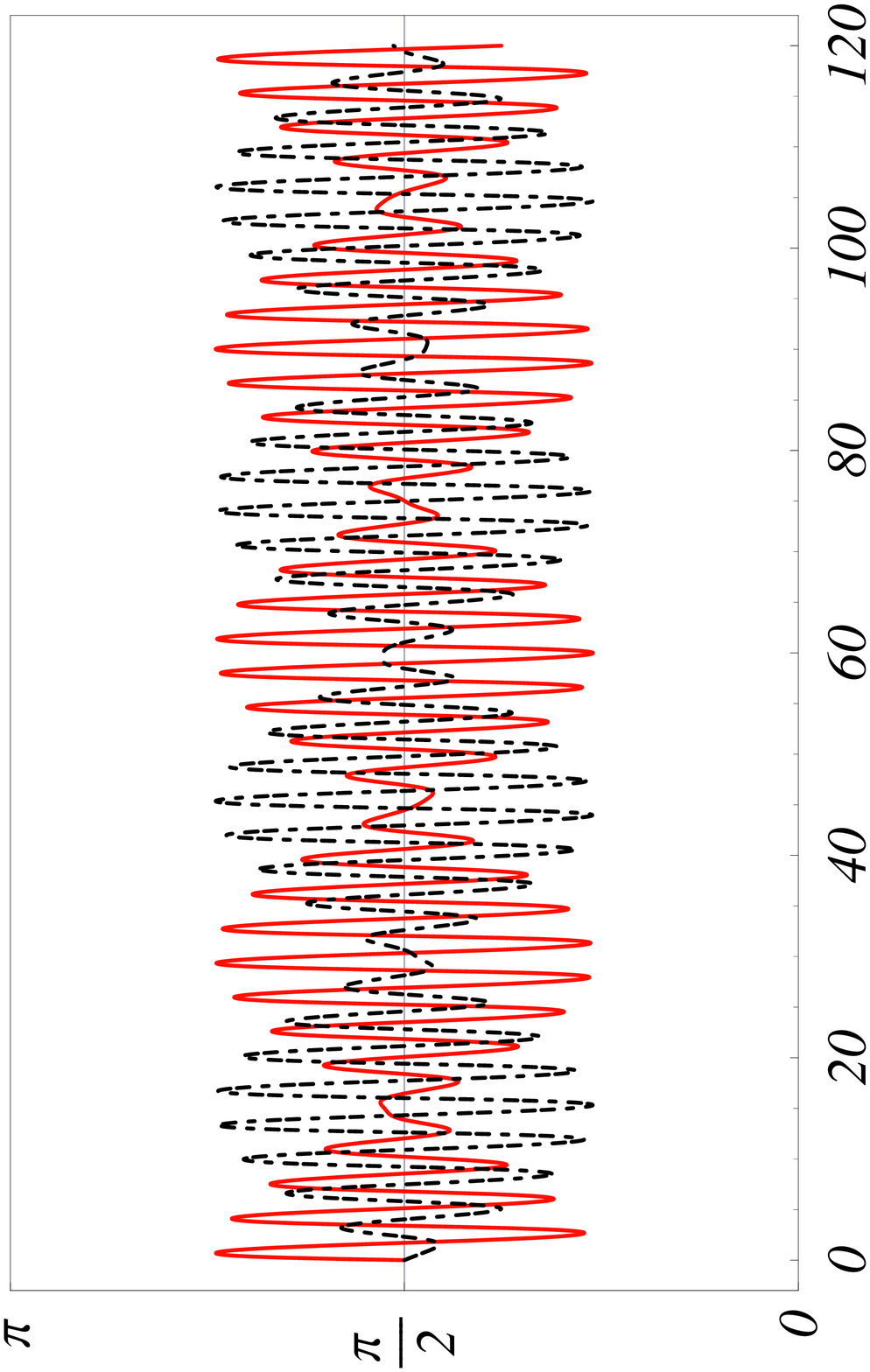}
\includegraphics[width=3cm,height=5.5cm,angle=270]{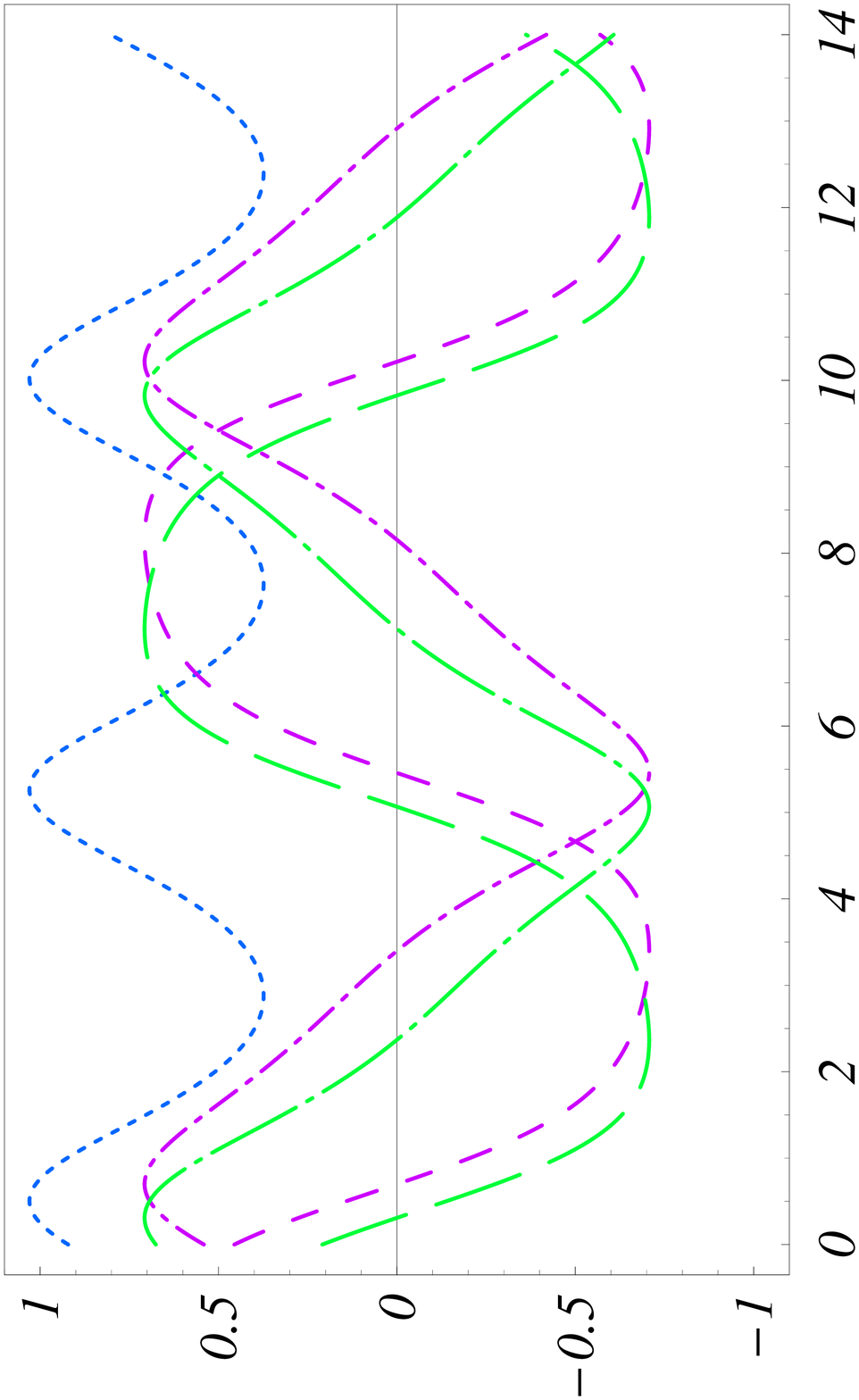}
\includegraphics[width=3cm,height=5.5cm,angle=270]{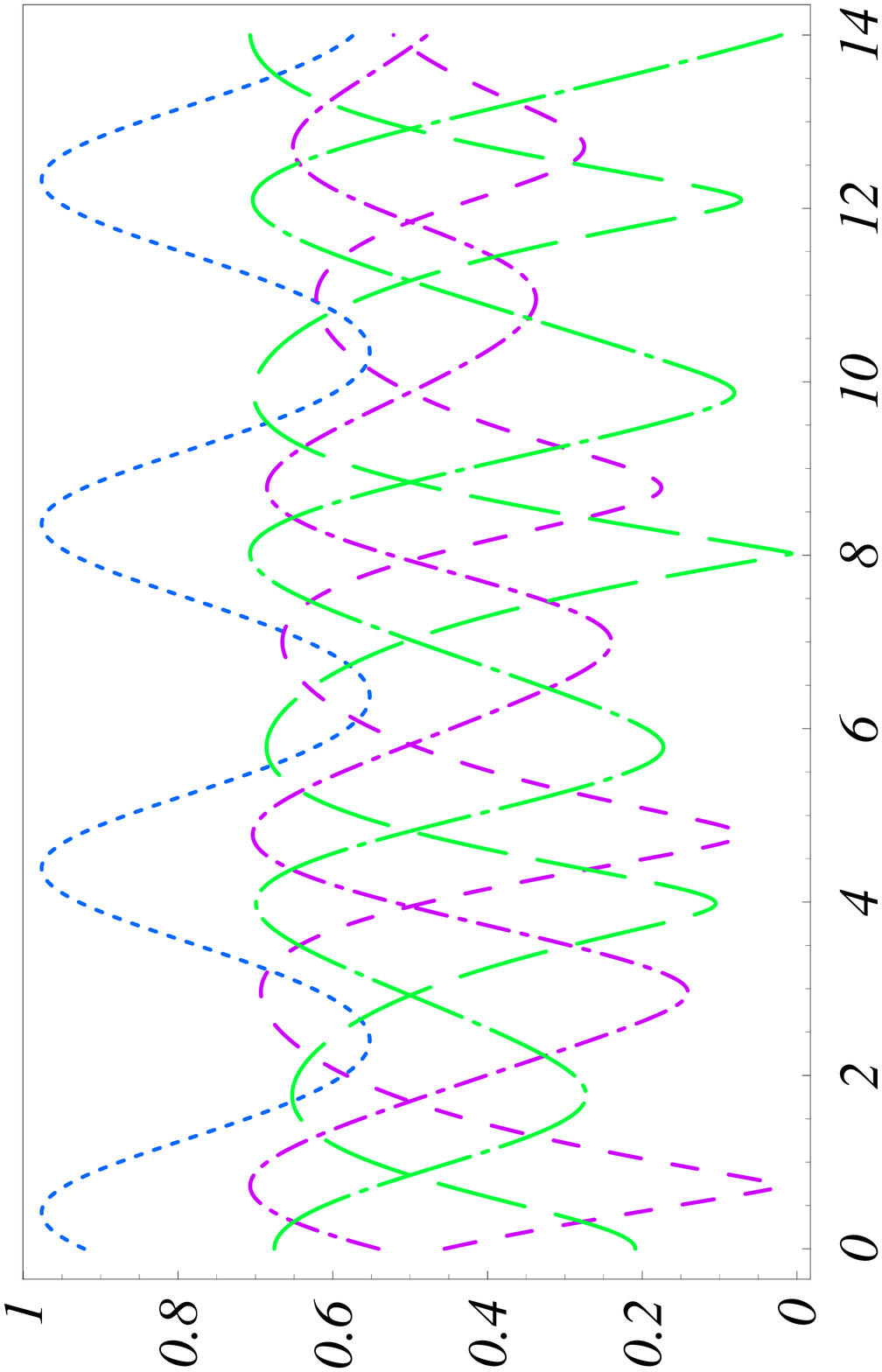}
\includegraphics[width=3cm,height=5.5cm,angle=270]{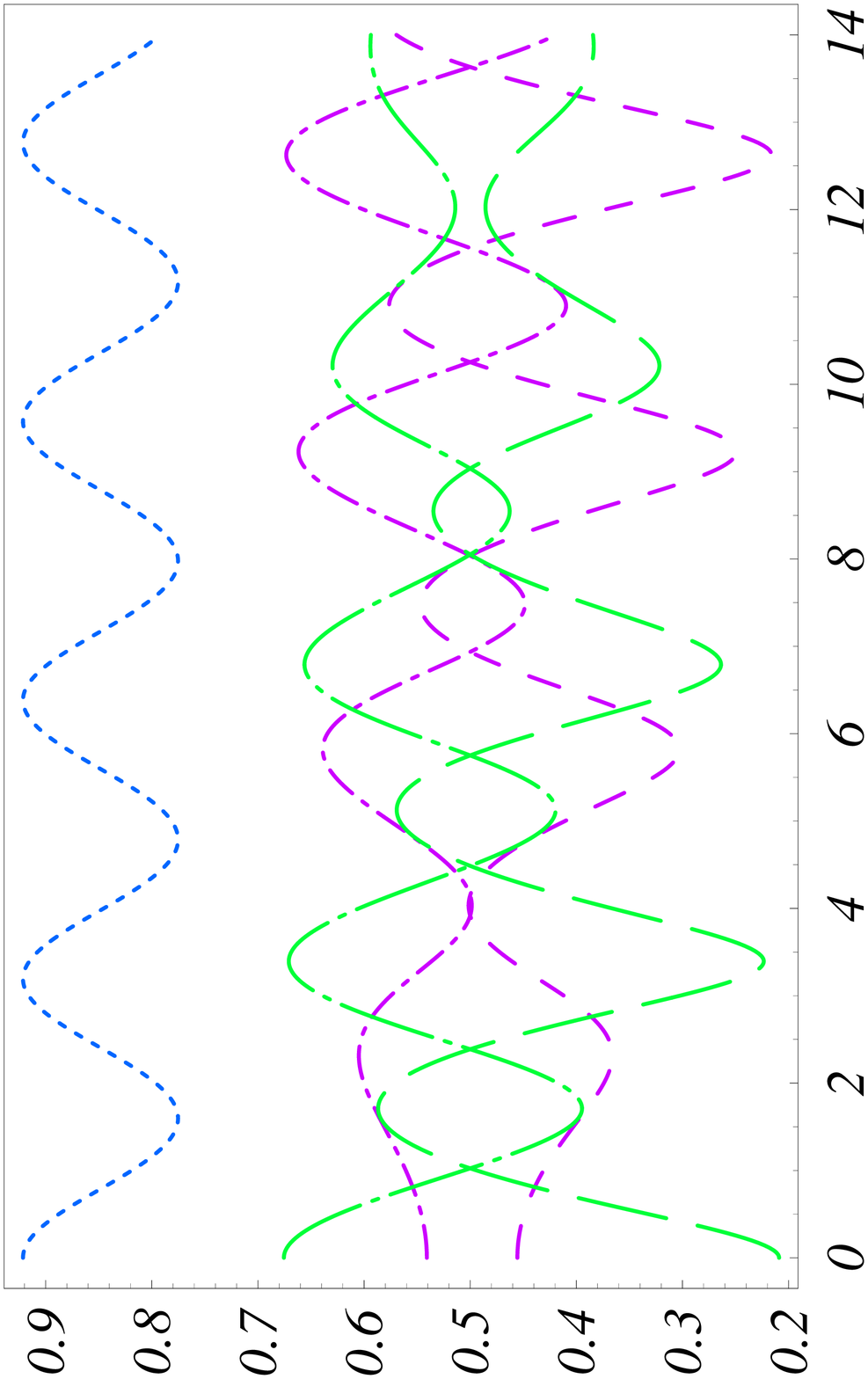}
\includegraphics[width=3cm,height=5.5cm,angle=270]{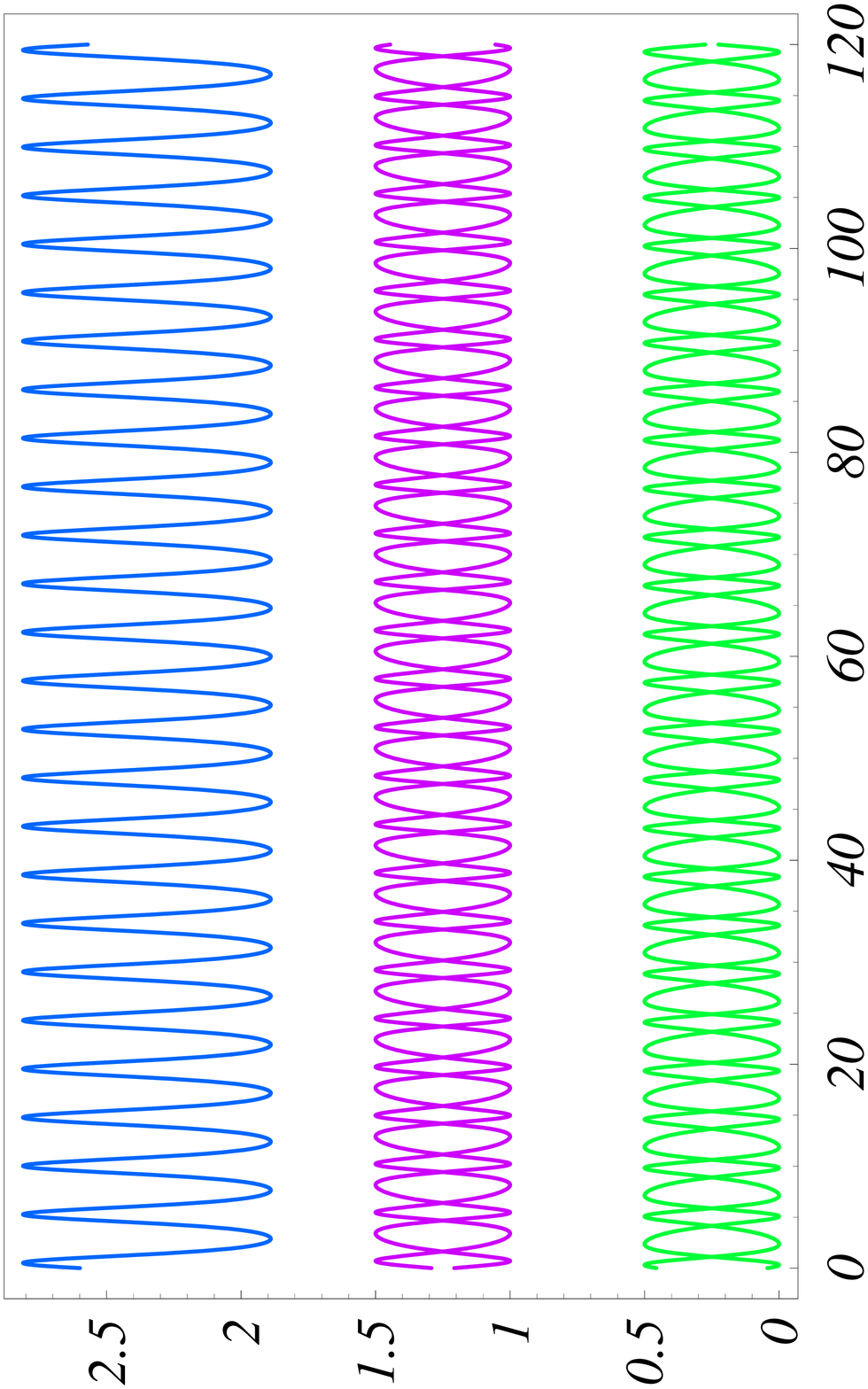}
\includegraphics[width=3cm,height=5.5cm,angle=270]{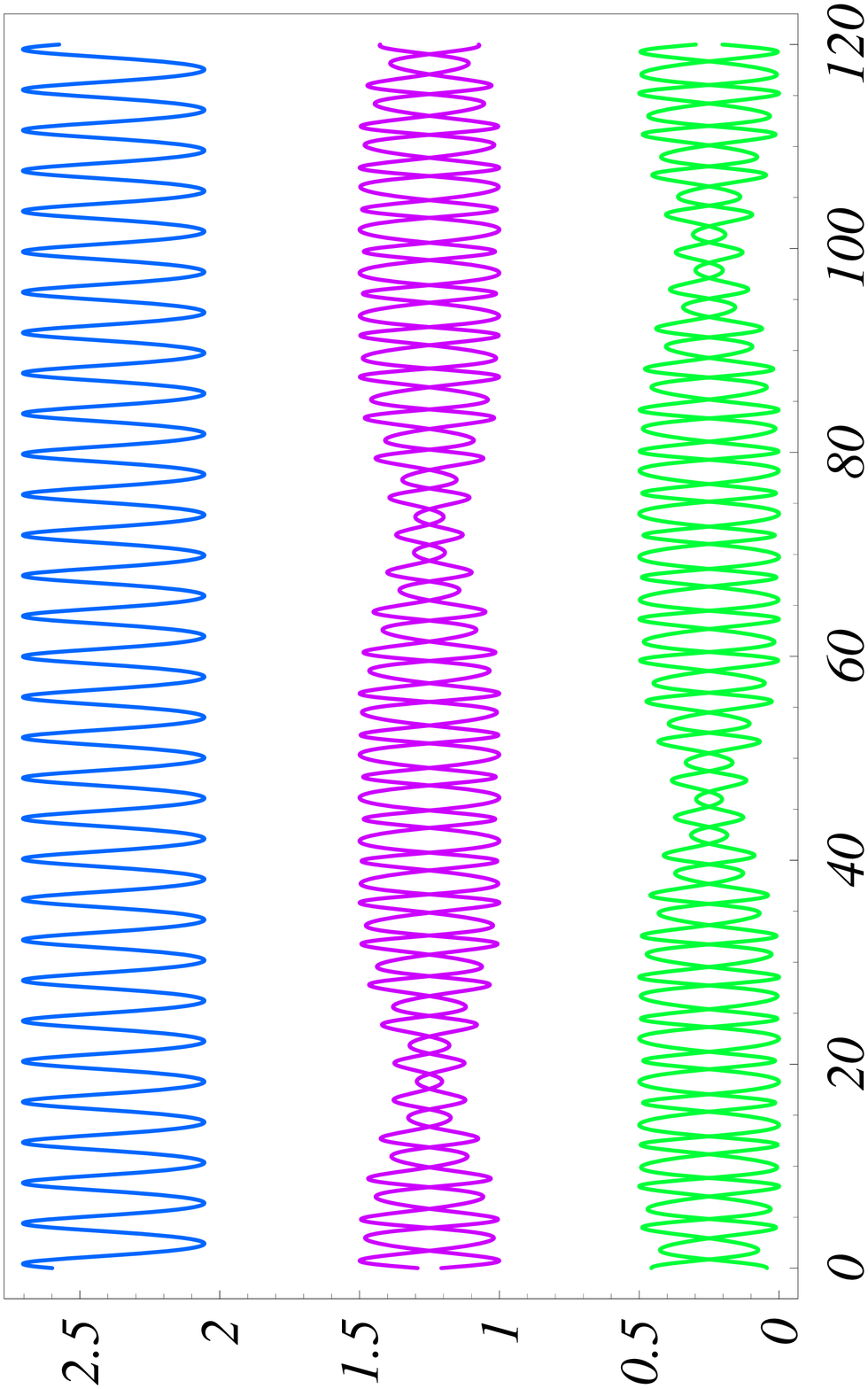}
\includegraphics[width=3cm,height=5.5cm,angle=270]{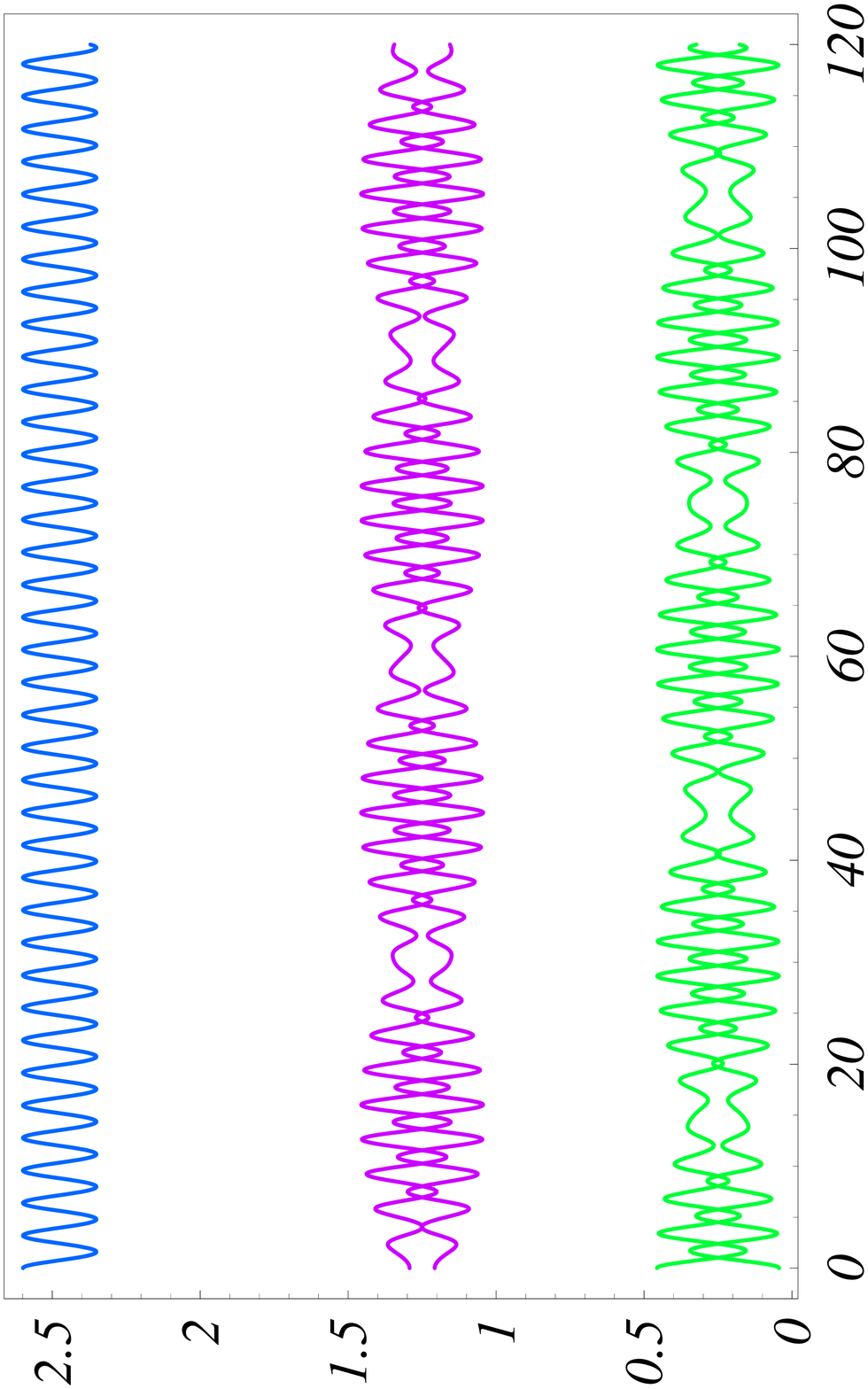}
\end{center}
\caption{\label{2} Color online. \textbf{Upper panel}. Plots of dynamical phases
as functions of time, for a butterfly with $Z_a=Z_b=1$. For each frame,
$\varphi_a(t)$ is (red) solid and $\varphi_b(t)$ is (black) dashed. Initial
conditions $\a_a=0.3, \a_b= 0.7;$ and conserved quantities $I_{ab}=1.1,
I_{23a}=I_{23b}=0.5 $ for all frames. Initial conditions for the phases and
corresponding cubic conserved quantities: \textbf{Left}: $\varphi_a=\varphi_b=0,
I_B=0.$ \textbf{Middle}: $\varphi_a=0, \varphi_b=\pi/2, I_B=0.13.$ \textbf{Right}:
$\varphi_a= \varphi_b=\pi/2, I_B=0.36.$ \textbf{Middle panel}. Plots of modes'
amplitudes of a butterfly as functions of time, same initial conditions and
parameters as in the corresponding left, middle and right frames of upper panel.
Connecting mode $C_1$ is (blue) dotted,
 $C_{2a}$ is (green) long-dash-dotted, $C_{3a}$ is (green) long-dashed,
 $C_{2b}$ is (purple) short-dash-dotted, $C_{3b}$ is (purple) short-dashed.
\textbf{Lower panel}. Plots of  amplitudes squared, as functions of time, same
initial conditions and parameters as in the corresponding left, middle and right
frames of upper and middle panels. Colors as above. To facilitate view, $C_{2b}^2,
C_{3b}^2$ are shifted upwards by the value $1$, and  $C_1^2$ is shifted upwards by
the value $1.75$. Here, in all three panels, the horizontal axe denotes
non-dimensional time; the vertical axe denotes phase in the upper panel and
amplitudes in the middle and lower panels.}
\end{figure*}

\noindent \textbf{3. Triad, case $I_T \neq 0$.}  Now we present some results for
the cases when $I_T \neq 0$ but otherwise the resonance conditions in the standard
form (\ref{res}) are satisfied. In Fig.\ref{1}, the characteristic evolution of
amplitudes is shown, depending on the value of $I_T$. To characterize the initial
conditions, the variable $\a=\arctan(C_3/C_2)$ has been chosen for a triad and
variables $\a_a=\arctan(C_{3a}/C_{2a})$ and $\a_b=\arctan(C_{3b}/C_{2b})$ have
been chosen for a butterfly. These variables appear naturally from the explicit
form of the corresponding Hamiltonians, and simplify the form of the dynamical
systems and conservation laws (see \cite{BK08} for more details).

\noindent  As it is shown in Fig.\ref{1}, when initial dynamical phase is zero it
will remain zero at all times, but the amplitudes will change sign periodically
(upper-left panel). When dynamical phase is initially very small but non-zero, the
amplitudes become purely positive and dynamical phase will have abrupt jumps, at
those times when the amplitudes used to change sign (lower-left panel).
Physically, in terms of squares of amplitudes, the dynamics in both cases is quite
the same (figure not shown). However, the phase's dynamics, with its periodic
motion, is revealed in the non-zero case.  As it is shown in upper panel (left,
middle and right), non-zero dynamical phase influences the evolution of
amplitudes, so that as initial $\varphi$ increases from $0$ to $\pi/2$, the range
of amplitude variations decreases from 1 to 0.1 and the period of the motions
decreases from 5 to 3.

\noindent In Fig.\ref{1}, lower panel,  we show that the notion of A-mode (active)
and P-mode (passive) introduced in \cite{KL-08} (compare to stability criterion
\cite{H67}) is useful also in the case of non-zero dynamical phase $\varphi$.
A-mode is the mode with the highest frequency, and two other modes are called
P-modes. In the pictures, $C_1$ is P-mode and $C_3$ is A-mode. Lower-middle
picture: when initial value of amplitude $C_1
>> C_3, C_2$ (4 times on figure), the P-mode $C_1$ keeps its energy
and the A-mode $C_3$ interacts strongly with the remaining P-mode $C_2$. If $C_2
>> C_3, C_1$ then the situation will be qualitatively the same, with the P-mode
$C_2$ keeping the energy. Lower-left figure: on the other hand, if $C_3 >> C_1,
C_2$, then a completely different time evolution is observed and all modes
interact.

One more important general feature of the dynamical phase is shown on the upper-
and lower-right panels. Indeed, independently of details of the initial values of
$C_1, C_2, C_3$, the variation range of the amplitudes is minimized when the
initial condition for the dynamical phase $\varphi$ is equal to $\pi/2$. This can
be used in real physical systems in order to control the exchange of energy
between resonant modes, \emph{at no energy cost}: the choice of initial dynamical
phase does not change the energy of the system, which is a sum of squares of
amplitudes, obviously independent of the dynamical phase (and of any phase, for
that matter).

\noindent \textbf{4. Butterfly with complex amplitudes.} As it was shown
\cite{KL-08}, clusters formed by two triads  $a$ and $ b $ connected \emph{via}
one mode can have one of three types accordingly to the types of connecting mode
in each triad: PP-, AP- and AA-butterfly. In this Letter, a PP-butterfly is taken
as a representative example. Dynamical system describing evolution of a
PP-butterfly
 has the form
 \bea \label{PP}
\begin{cases}
\dot{B}_{1}=  Z_a B_{2a}^*B_{3a} +   Z_b B_{2b}^*B_{3b}\,, \\
\dot{B}_{2a}=  Z_a B_{1}^* B_{3a}\,,\quad    \dot{B}_{2b}=  Z_b B_{1}^* B_{3b}\,, \\
\dot{B}_{3a}=  - Z_{a} B_{1} B_{2a} \,,\    \dot{B}_{3b}=  - Z_{b} B_{1} B_{2b}\ . \\
\end{cases}
 \eea%
where notation $B_{1} = B_{1a}=B_{1b}$ is chosen for the amplitude of the mode
common for both triads while $B_{2a}, B_{3a}, B_{2b}, B_{3b}$ are other four modes
of the butterfly cluster.
The set of constructed conservation laws reads as%
 \bea \label{int-PP}
\begin{cases} I_{23a}=|B_{2a} |^2 + |B_{3a}|^2 \,, \quad
 I_{23b}= |B_{2b} |^2 + |B_{3b}|^2 \,, \\
 I_{ab}= |B_{1}|^2 + |B_{3a} |^2 + |B_{3b}|^2 \,, \\
I_B = \operatorname{Im}(Z_a B_1 B_{2a} B_{3a}^* + Z_b B_1 B_{2b} B_{3b}^*)
 \end{cases}
  \eea
Similar to the triad, the use of standard representation $B_j= C_j\exp(i
\theta_j)$ shows that the Sys.(\ref{int-PP}) has effectively 3 degrees of freedom
and two dynamical phases are important: $ \varphi_{a} = \theta_{1a} + \theta_{2a}
-\theta_{3a}, \quad \varphi_{b} = \theta_{1b} + \theta_{2b} -\theta_{3b}, $
 with  the requirement
$\theta_{1a}=\theta_{1b}$ which corresponds to the choice of connecting mode.
Accordingly,
 equations on the dynamical phases take form%
\bea \label{DynPhaseB}
\begin{cases}
\dot{\varphi}_a= -I_B (C_1^{-2}+C_{2a}^{-2}-C_{3a}^{-2}), \\
\dot{\varphi}_b= -I_B (C_1^{-2}+C_{2b}^{-2}-C_{3b}^{-2}).
\end{cases}
 \eea

\noindent In order to study the effect of non-zero dynamical phases for butterfly,
we performed numerical simulations for the integrable case $Z_a = Z_b$. This
allows us to compare the results with the triad, which  is just a particular case
of the integrable butterfly.

In Fig.\ref{2}, left column: we show phases, amplitudes and amplitudes squared
(energies) for initial conditions $\varphi_{a}=\varphi_{b}=0$. The dynamics is
quite similar to the triad with initial condition $\varphi = 0$ shown in
Fig.\ref{1} upper left. In Fig.\ref{2}, middle column: phases, amplitudes and
energies are shown for initial conditions $\varphi_{a}=\pi/2, \varphi_{b}=0$,
while in the right column the same data are presented, for initial conditions
$\varphi_{a}=\varphi_{b}=\pi/2.$
We observe from Figs.\ref{1} and \ref{2} some effects from the dynamical phases
$\varphi_{a}(t)$ and $\varphi_{b}(t)$ of a butterfly (correspondingly, the phase
$\varphi(t)$ of a triad): to diminish the period of energy exchange $\tau$ within
a cluster by 20$\%$ and more; to reduce the variability of wave energies by 25$\%$
and more; to generate a new time scale, $T >> \tau$, in which there is
considerable energy exchange within a cluster, as well as a periodic behaviour
(with period $T$) in the variability of modes' energies.

All computations have been done using Mathematica and we have validated the code
by checking the corresponding conservation laws, particularly those of cubic and
quartic dependence on the amplitudes (introduced in \cite{BK08}). These
conservation laws are stably conserved during the whole numerical simulation,
within a relative error of $10^{-12}$.

A comment regarding the ergodicity of the integrable butterfly. We observe in a
parametric plot of $\cos(\varphi_a)$ vs. $\cos(\varphi_b)$ as functions of time
(figure not shown), that the seemingly periodic motions are indeed precessing with
precession speed depending on the initial conditions. This is a generic feature of
integrable systems which are not superintegrable.

\noindent \textbf{5. Conclusions.} Effects of non-zero dynamical phases should be
taken into account in the following situations.

\textbullet~ To control energy input, exchange and output in
laboratory experiments, e.g. in Tokamaks. Indeed, a possibility of
concentrating energy in a small set of drift waves via some
instability mechanisms has been conjectured by V.I.Petviashvili some
15 years ago \cite{Pet}. As soon as energy is concentrated in a
resonant cluster, mode amplitudes can become dangerously large. In
\cite{Hmode} it was shown that the appearance of resonances can be
completely avoided by special choice of the form of the laboratory
facilities which, of course, is too costly a game with Tokamaks. On
the other hand, adjustment of dynamical phases can diminish
amplitudes of resonantly interacting drift waves 10 times and more
for the same technical equipment.

\textbullet~ To gain more insight into the phenomenon of zonal flows in plasmas
which are now regarded as the main component in all regimes of drift wave
turbulence. ``The progress of plasma physics induced a paradigm shift from the
previous `linear, local and deterministic' view of turbulent transport to the new
`nonlinear, nonlocal (both in real and wave number space), statistical' view of
turbulent transport. Physics of the drift wave-zonal flow system is a prototypical
example of this evolution in understanding the turbulence and structure formation
in plasmas''(\cite{Diamond}). In \cite{Con_et_al}, a modulational instability of
Rossby/plasma drift waves leads to generation of zonal jets through a process in
which the wave amplitudes are initially well-approximated by a \emph{kite}, a
cluster consisting of two triads connected \emph{via} two modes. The study of the
behaviour of associated dynamical phases could lead to a deeper understanding of
zonal jet formation. Structure formation (for 3-wave resonance processes) is
presented in \cite{KM07}, examples of non-local interactions are given in
\cite{AMS} as well as the cases of `weak' locality (waves with wave numbers of
order $n$ and $n^2$ can form a resonance cluster,  $n$ and $n^3$ - can not);
effects of initial energy distribution among the modes of a cluster are studied in
\cite{KL-08}. In this Letter we identify dynamical phase as an additional
important parameter for any theoretical study of nonlinear wave systems. We would
like to point out that the explicit equation for the dynamical phase of a triad
has been known for more than 40 years (Eq. (3.31), \cite{tsyt}) in plasma physics
and probably even earlier in nonlinear optics. We consider as our material impact
in this metier the detailed study of dynamical phases' effect on nonlinear
evolution of a triad and a butterfly.

 \textbullet~ To guide and improve the performance and analysis of laboratory
experiments. We showed that non-zero phases can dramatically reduce the
variability of the oscillations (Figs. \ref{1} and \ref{2}, left columns). It
would then be possible to tune initial conditions and/or forcing (in rotating
water tanks, for example) in order that the measurement of the resonant modes'
oscillations be less subject to errors. In \cite{CHS96} results of laboratory
experiments with gravity-capillary waves are presented, and corresponding
dynamical system for three connected triads is written out and solved explicitly.
The authors report qualitative agreement of the observation with the solutions of
the dynamical system though the magnitudes of observed amplitudes are higher than
those theoretically predicted. For all calculations the dynamical phase of the
initially excited triad was set to $\pi$ which might be the source of this
discrepancy.

\textbullet~ To interpret the results of numerical simulations. For instance, in
\cite{KL-06} a generic model of intra-seasonal oscillations in the Earth's
atmosphere has been presented which describes the processes with periods of the
order of 30-90 days. In this model, dynamical phase has not been taken into
account yet. A new time scale, $T >> \tau$,  which is clearly observable in Fig.
\ref{2}, corresponds, for the resonant triads of atmospheric planetary waves, to
periods of the order of 2-5 years. This is the time range of climate variability.
This means that in numerical modeling of the climate variability the control of
dynamical phase is a matter of prodigious importance.

\noindent {\bf Acknowledgements}.  The authors wish to thank the
organizing committee and participants of the workshop ``INTEGRABLE
SYSTEMS AND THE TRANSITION TO CHAOS II'' for  stimulating
discussions. The authors also highly appreciate the hospitality of
Centro Internacional de Ciencias (Cuernavaca, Mexico), where part of
this work was completed. E.K. acknowledges the support of the
Austrian Science Foundation (FWF) under project P20164-N18. M.B.
wishes to thank the support of the organizers of the Programme ``The
Nature of High Reynolds Number Turbulence'' at Isaac Newton
Institute, Cambridge and many participants, including P. Bartello,
C. Cambon, C. Connaughton, N. Grisouard, M. McIntyre, K. Moffatt, S.
Nazarenko, J. J. Riley, J. Sommeria and C. Staquet, for helpful
discussions. M.B. acknowledges the support of the Transnational
Access Programme at RISC-Linz, funded by 6 EU Programme SCIEnce
(Contract No. 026133).

\end{document}